# NuSeC: A Dataset for Nuclei Segmentation in Breast Cancer Histopathology Images

**Organizers**

- Department of Computer Engineering, Ankara University, Ankara, Turkey
    - Prof. Dr. Refik Samet, samet@eng.ankara.edu.tr
    - PhD Student Zeynep Yildirim, yildirimz@ankara.edu.tr
    - PhD Student Nooshin Nemati, nntolakan@ankara.edu.tr
    - MSc Student Mohamed Traore, mtraore@ankara.edu.tr

- Department of Software Engineering, Mehmet Akif Ersoy University, Burdur, Turkey
    - Assoc. Prof. Dr. Emrah Hancer, ehancer@mehmetakif.edu.tr
- Department of Medical Pathology, Ankara University, Ankara, Turkey
    - Prof. Dr. Serpil Sak, sak@medicine.ankara.edu.tr
    - Assoc. Prof. Dr. Bilge Ayca Kirmizi, akarabork@yahoo.com

## 1 Introduction

Breast cancer is the most frequently diagnosed form of cancer and is the second leading cause of death caused by cancer in women. It is necessary to ensure patient survival with a multidisciplinary approach, including patholo gists, medical, surgical, and radiation oncologists. In order to diagnose breast cancer type, stage, and grade accurately, examination of tissue biopsies and operation specimens is necessary. The gold-standard clinical diagnosis of cancer relies on the examination of biopsy specimens prepared onto slides. The biopsy specimens must be fixed embedded in paraffin blocks, mounted on glass slides and stained. Hematoxylin and Eosin (H&E), is a routine stain used in pathology laboratories all over the globe, which gives a good contrast of a tissue section and is commonly used to identify nuclei and cytoplasm [1]. Nevertheless, histopathological examination of the prapared slides involves laborious, time-consuming processes that are limited by specimen quality and pathologist experience. Therefore, pathological workflows are being automated and digitized, and computer-aided diagnosis (CAD) is being used to streamline and optimize tissue analysis [2]. The goal of this study is to introduce a publicly available dataset (called NuSeC) for the segmentation process of nuclei in H&E stained breast cancer histopathology images. It is therefore expected to develop robust and reliable cancer diagnosis CAD systems with the help of the introduced NuSeC dataset. The dataset is created through H&E stained breast slides of 25 different invasive breast carcinoma no special type (NST) patients captured at 40x magnification from the Department of Medical Pathology at Ankara University. The slides have been scanned by 3D Histech Panoramic p250 Flash-3 scanner and Olympus BX50 microscope. Furthermore, the annotation process of nuclei structures from H&E stained images is manually carried out by using the QuPath software. Accordingly, a mask image is created for each image.

## 2 Training/Testing Datasets

The NuSeC dataset is created by selecting 4 images with the size of 1024×1024 pixels from the slides of each patient among 25 patients. Therefore, there are a total of 100 images in the NuSeC dataset. To carry out a consistent comparative analysis between the methods that will be developed using the NuSeC dataset by the researchers in the future, we divide the NuSeC dataset 75% as the training set and 25% as the testing set. In detail, an image is randomly selected from 4 images of each patient among 25 patients to build the testing set, and then the remaining images are reserved for the training set. While the training set includes 75 images with around 30000 nuclei structures, the testing set includes 25 images with around 6000 nuclei structures. Some image examples and their annotations from the NuSeC dataset are presented in Fig. 1.

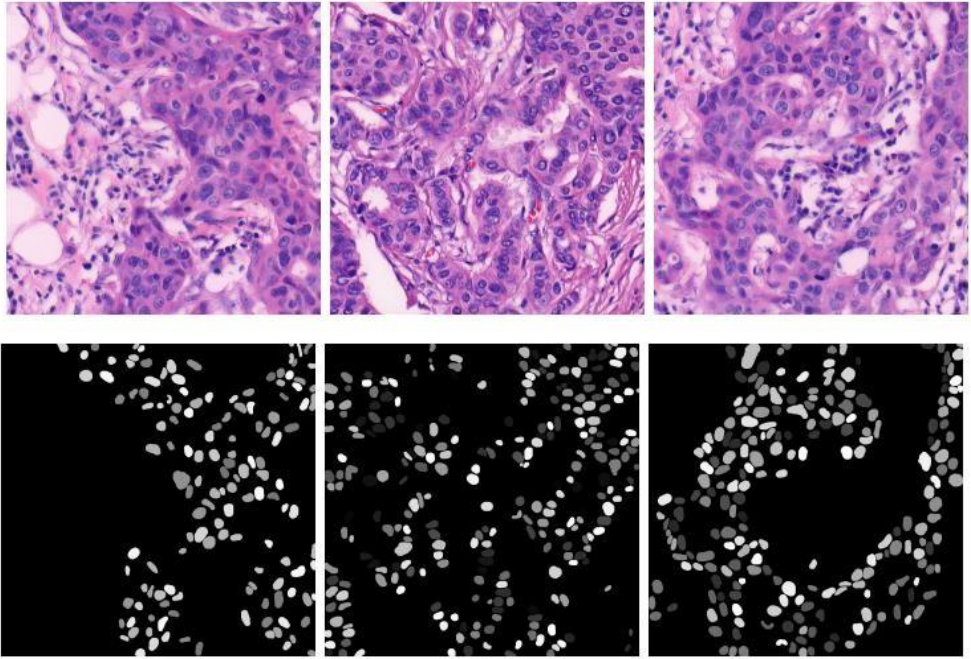

Figure 1: First and second rows represent some original H&E images and their annotation masks respectively.

## 3 Evaluation

The metrics to evaluate the goodness of a algorithm on the NuSeC dataset are as follows.

  *a) Aggregated Jaccard Index (AJI):* The AJI metric [3] is one of the recently introduced index to evaluate the nuclei segmentation performance. The value of AJI ranges between 0 to 1 (higher is better). Let $G = \{G_i\}_{i=1}^{N}$ be the ground truths segments in an image, $N$ denotes the total amount of segments in $G$; and let $S = \{S_k\}_{k=1}^{M}$ be the predicted segments of the corresponding image and $M$ denotes the total amount of segments in $S$. AJI is defined as follows.

$$AJI = \frac{\sum_{i=1}^{N} G_i \cap S_i}{\sum_{i=1}^{N} G_i \cup S_i + \sum_{s_{k \in U}} S_k},$$

where $S_i$ is the matched predicted segments that produce the largest Jaccard Index value with $G_i$; and U denotes the set of unmatched predicted segments, where the total amount of U is *(M-N)* [4].

  *b) Intersection over Union (IoU):* IoU is one of the most commonly used metric. IoU measures the intersection over the union of nuclei areas on a ground truth mask. IoU is defined as follows.

$$IoU = \frac{TP}{TP+FP+FN},$$

where TP, FP and FN are respectively true positives, false positives and false negatives.

**NuSeC Dataset Access Link**

The NuSeC dataset can be downloaded from the following link:
Dataset available here

**Acknowledgement**

This work is supported by Turkish Scientific and Research Council (TUBITAK) under Grant No.121E379.